\documentclass{aa}  

\usepackage{subcaption} 
\usepackage{amsmath}    
\usepackage{float}     
\renewcommand{\linenumbers}{}  
\usepackage{graphicx}
\usepackage{gensymb}
\usepackage[T1]{fontenc}
\usepackage{amssymb}
\usepackage{natbib}
\usepackage{tabularx}
\usepackage[utf8]{inputenc}
\usepackage[english]{babel}

\usepackage{txfonts}
\usepackage{ulem}
\usepackage{xcolor}

\usepackage{hyperref}

\DeclareUnicodeCharacter{02BC}{'}

\begin{document}

\title{A nearby FR I type radio galaxy 3C 120 as a possible PeV neutrino emitter}

\author{
Rong-Qing Chen\inst{\ref{gzdx}}
\and
Neng-Hui Liao\inst{\ref{gzdx} \thanks{$*$ Corresponding authors: nhliao@gzu.edu.cn, yzfan@pmo.ac.cn}} \thanks{\href{https://orcid.org/0000-0001-6614-3344}{ORCID: 0000-0001-6614-3344}}
\and
Xiong Jiang\inst{\ref{zjstwt},\ref{zkd}}
\and
Yi-Zhong Fan\inst{\ref{zjstwt},\ref{zkd}}
\thanks{\href{https://orcid.org/0000-0002-8966-6911}{ORCID: 0000-0002-8966-6911}}
}

\institute{
Department of Physics and Astronomy, College of Physics, Guizhou University, Guiyang 550025, Peopleʼs Republic of China\label{gzdx}
\and
Key Laboratory of Dark Matter and Space Astronomy, Purple Mountain Observatory, Chinese Academy of Sciences,
Nanjing 210023, People’s Republic of China\label{zjstwt}
\and
School of Astronomy and Space Science, University of Science and Technology of China, Hefei, Anhui 230026, People’s Republic of China\label{zkd}
}
\date{Received XXX; accepted YYY}

\abstract
	{Although connections between flaring blazars and some IceCube neutrinos have been established, the dominant sources for the bulk extragalactic neutrino emissions are still unclear and one widely suggested candidate is a population of radio galaxies. Because of their relatively low $\gamma$-ray radiation luminosities ($L_\gamma$), it is rather challenging to confirm such a hypothesis with the neutrino/GeV $\gamma$-ray flare association. Here we report on the search for the GeV $\gamma$-ray counterpart of the neutrino IC-180213A and show that the nearby ($z$ = 0.03) broad line radio galaxy 3C 120 is the only  known co-spatial GeV $\gamma$-ray source in a half-year epoch around the neutrino detection. An intense $\gamma$-ray flare, the second strongest one among the entire 16-year period, is temporally coincident with the detection of IC-180213A. Moreover, accompanying optical brightenings in $g$-band and $V$-band are observed. We also find that the IC-180213A / 3C 120 association follows the $L_\gamma$-$D_{L}^{2}$ correlation for the neutrino sources (candidates), including NGC 1068 and some blazars. These facts suggest that 3C 120 is a candidate for emitting high-energy neutrinos and may offer an initial evidence for the radio galaxy origin of some PeV neutrinos.
}

\keywords{
	neutrino astronomy 
        --
	active galactic nuclei 
	-- 
	Gamma-ray sources
	}
 \titlerunning{3C 120 as a possible PeV neutrino emitter}
\authorrunning{Chen et al.}
\maketitle

\section{Introduction}
\label{sec:Introduction}

Neutrinos, which interact weakly with matter, can escape from extreme astrophysical environments that are impenetrable to electromagnetic radiation and hence provide unique insights into these intriguing astrophysical environments. The IceCube neutrino observatory at the South Pole\footnote{http://icecube.wisc.edu} has discovered some high-energy neutrinos of astrophysical origin \citep[e.g.,][]{2013Sci...342E...1I,2015PhRvL.115h1102A,2023Sci...380.1338I}.
These neutrinos were produced through the interaction of ultra-high energy cosmic rays (UHECRs) with ambient matter (i.e., $p$-$p$) or radiation fields (i.e., $p$-$\gamma$), hence are crucial for revealing the origin of the UHECRs. The production of such neutrinos would generate electromagnetic radiations, in particular $\gamma$-ray photons, as well. The proposed $\gamma$-ray emitters for extragalactic high energy neutrinos include the blazars \citep{2001PhRvL..87v1102A}, starburst galaxies \citep{2006JCAP...05..003L}, radio galaxies \citep{2014PhRvD..89l3005B}, and galaxy clusters \citep{2008ApJ...689L.105M}. The neutrino/$\gamma$-ray association is a powerful tool to probe the origin of the IceCube events. So far, the blazar TXS 0506+056 as well as the very nearby Seyfert II galaxy NGC 1068 have been confirmed to be the sources of the TeV-PeV neutrinos \citep{2018Sci...361.1378I,2022Sci...378..538I}.

The advantage of establishment of a connection between a flaring blazar and an incoming neutrino is that both the spatial and temporal information can be utilized. Cases similar with TXS 0506+056/IceCube-170922A are mounting  \citep[e.g.,][]{2016NatPh..12..807K,2019ApJ...880..103G,2020A&A...640L...4G,2020ApJ...893..162F,2022ApJ...932L..25L,2023MNRAS.519.1396S,2024ApJ...965L...2J}. However, the fraction of contribution from blazars to the total observed neutrino flux is proposed to be limited, since no significant cumulative neutrino excesses are found from the Fermi blazar directions \citep{2017ApJ...835...45A}.

The jets of active galactic nuclei (hereafter jetted AGNs) are the dominant population of the extragalactic $\gamma$-ray sky, in which majority of them are blazars \citep{2023arXiv230712546B}. Benefited from the strongly Doppler boosted effect due to the well aligned relativistic jets, the broadband non-thermal emissions of blazars are overwhelming and highly variable \citep{1978bllo.conf..328B,1997ARA&A..35..445U,2019ARA&A..57..467B}. By comparison, radio-loud AGNs with misaligned jets (i.e., MAGN), including
radio galaxies and steep spectrum radio quasars, are characterized by steep radio spectra as well as bipolar radio structures \citep{1993ARA&A..31..473A,1995PASP..107..803U}. Although their radiations are mildly Doppler boosted, MAGNs are prominent GeV $\gamma$-ray emitters \citep{2010ApJ...720..912A,2022ApJS..263...24A}. A few MAGNs are even shining in the TeV $\gamma$-ray sky  \citep[e.g.,][]{2003A&A...403L...1A,2009ApJ...695L..40A,2012A&A...539L...2A}. Nearby radio galaxies have been considered as promising UHECR emitters \citep{2022Univ....8..607R}. 

Recently, the IceCube Event Catalog of Alert Tracks (i.e., ICECAT-1) has been released \citep{2023ApJS..269...25A}. A 0.111 PeV track-like neutrino event (i.e., IC-180213A, a so-called bronze event), in an arrival direction event of R.A. $66.97^{+2.46}_{-2.59} \, ^\circ$and decl. $6.09^{+1.95}_{-1.72} \, ^\circ$\footnote{Here a simple \href{https://dataverse.harvard.edu/dataset.xhtml?persistentId=doi:10.7910/DVN/SCRUCD}{ellipse} is adopted, shown in Figure \ref{tsmaps}.} (at a 90\% confidence level), is reported in ICECAT-1. Interestingly, a nearby Fanaroff-Riley (FR, \citealt{1974MNRAS.167P..31F}) I broad line radio galaxy (BLRG) 3C 120 falls into the localization uncertainty region of the neutrino. It is a well-studied nearby source with z = 0.033 \citep{1967ApJ...149L..51B}, $\simeq$ 145~Mpc away, hosting an efficiently accreting black hole with a mass of $5.5\times 10^{7} M_{\sun}$ by the reverberation mapping approach \citep{2004ApJ...613..682P}. Radio observations with the Very Long Baseline Array reveal super-luminal components with a median apparent speed of $\simeq$ 4.7c \citep{2021ApJ...923...30L}, and the jet inclination angle to the line of sight is constrained to be $\sim 10 - 20^\circ$. Consequently, a Doppler factor of $\delta \sim$ 2.4 and a bulk Lorentz factor of $\Gamma \sim$ 5 are suggested \citep{2005AJ....130.1418J,2009A&A...494..527H}. 3C 120 is famous for the strong variability from radio to GeV $\gamma$ rays, with which an accretion-disk-jet connection has been established \citep{2002Natur.417..625M,2010ApJ...720..912A,2011ApJ...740...29K}. In this paper, we thoroughly investigate the multi-wavelength data of 3C 120, to explore its potential association with the neutrino event IC-180213A. Data analyses are presented in Section 2; while discussions and a short summary are given in Section 3. We adopt a $\Lambda$CDM cosmology with $ \Omega_{M} $ = 0.32, $ \Omega_{\Lambda} $ = 0.68, and a Hubble constant of $H_{0}$ = 67 km$^{-1}$ s$^{-1}$ Mpc$^{-1}$ \citep{2014A&A...571A..16P}.

\section{Data analysis and results} \label{sec:2}

\subsection{Fermi-LAT Data} \label{subsec:2-1}

We collected the first 16-year (i.e., MJD 54683 - 60443, or from 2008 August 4 to 2024 August 4) Fermi-LAT Pass 8 SOURCE data (evclass = 128 and evtype = 3). Energy range of the data was selected between 100 MeV and 500 GeV. The Fermitools software version 2.2.0 was adopted for the data analysis, along with Fermitools-data version 0.18. In the initial data filtering procedure, a zenith angle cut (i.e., $<$ 90$ ^{\circ}$) was set to avoid significant contamination from the earth limb; meanwhile, the recommended quality-filter cuts (i.e., DATA\_QUAL==1 \&\& LAT\_CONFIG==1) were applied. Unbinned-likelihood analysis was performed by the gtlike task to extract the $\gamma$-ray flux and spectrum. We used the test statistic (TS = -2ln ($ L_{0}$/$L$), \citealt{1996ApJ...461..396M}) to determine the significance in the $\gamma$-ray detection. $L$ and $L_0$ correspond to the maximum likelihood values for the models with and without the target $\gamma$-ray source, respectively. During the likelihood analysis, a region of interest (ROI) of 10 degrees centered on the coordinates of 4FGL J0433.0+0522, for which the low-energy counterpart is 3C 120, was set. Parameters of the 4FGL-DR4 background source \citep{2023arXiv230712546B} within the ROI, as well as those for the diffuse emission templates (i.e., gll\_iem\_v07.fits and iso\_P8R3\_SOURCE\_V3\_v01.txt) were left free, while other parameters were frozen as the default values. In addition, the residual TS maps were produced in which potential $\gamma$-ray sources not included in the 4FGL-DR4 were checked. The model was updated by embracing these sources and the likelihood analyses were then re-performed. In the temporal analysis, weak background sources with TS $<$ 10 were removed from the model. Meanwhile, 95\% CL upper limits were obtained by the pyLikelihood UpperLimits tool, to replace the flux estimations.

An analysis of the entire 16-year dataset proves that 4FGL J0433.0+0522 is a significant $\gamma$-ray source (TS $\simeq$ 1500, 38$\sigma$). Its averaged flux is estimated as (4.9 $\pm$ 0.3) $\times$ 10$ ^{-8}$ ph cm$^{-2}$ s${-1}$, consistent with the results listed in 4FGL-DR4 \citep{2023arXiv230712546B}. It is confirmed as a spectrally soft $\gamma$-ray source (Spectral index $\simeq$ 2.83). Our localization analysis also confirms the spatial association between 3C 120 and the $\gamma$-ray source.

\begin{figure}[ht!]
\centering
\begin{subfigure}{0.85\columnwidth}
    \includegraphics[width=\columnwidth]{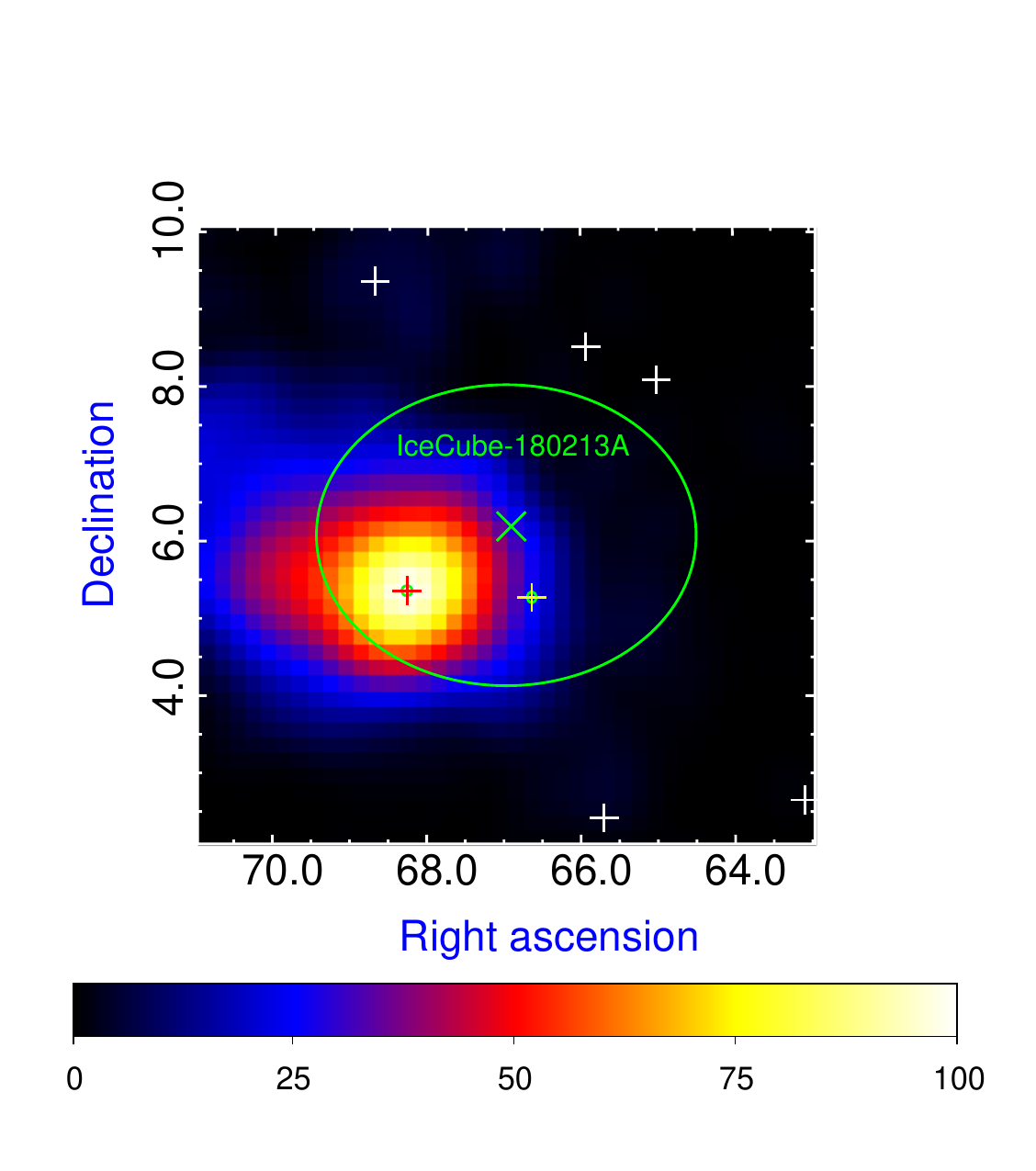}
    \caption{}
    \label{tsmaps}
\end{subfigure}
\begin{subfigure}{0.85\columnwidth}
    \includegraphics[width=\columnwidth]{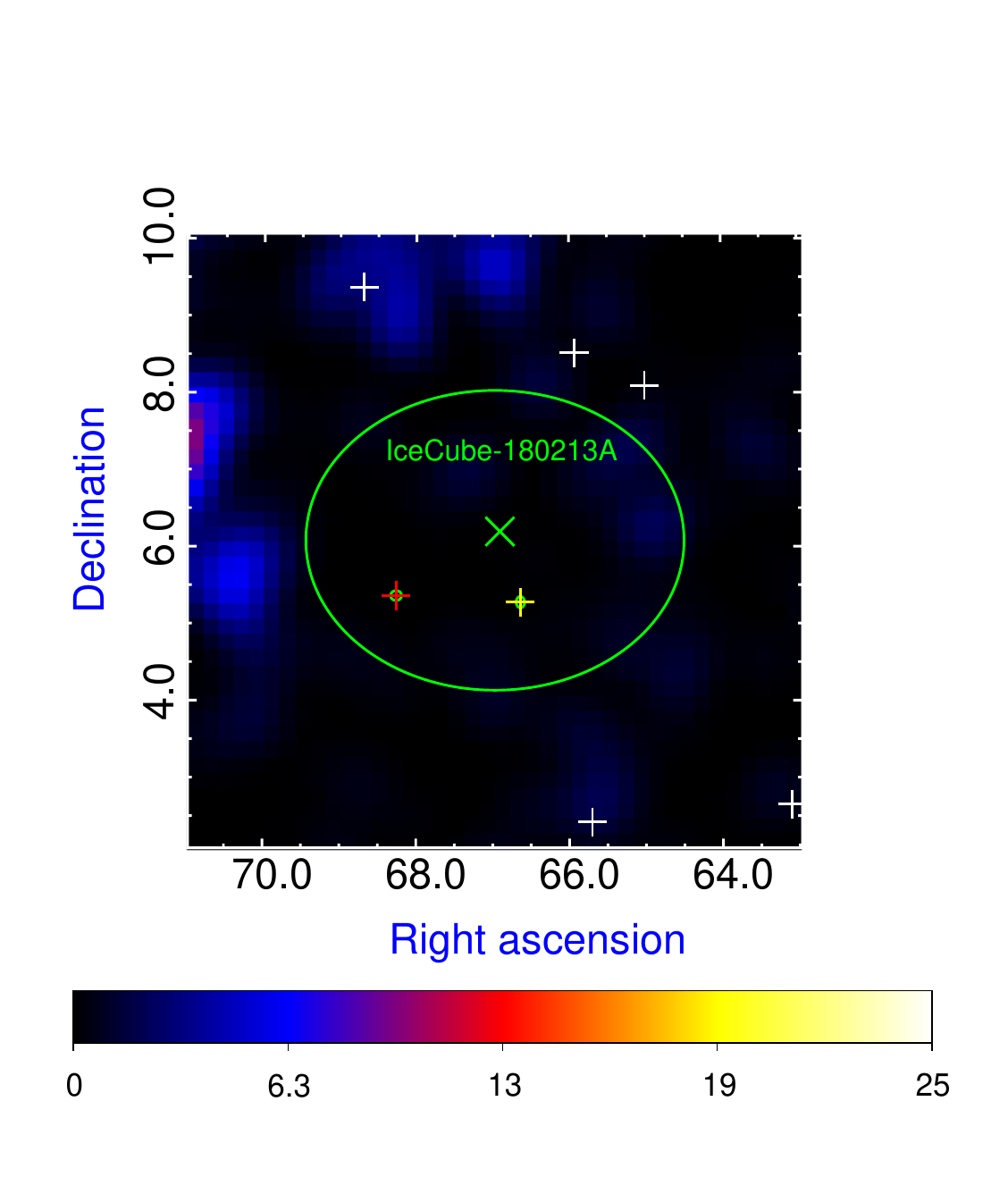}
    \caption{}
    \label{tsmapr}
\end{subfigure}
\caption{
 Smoothed $\gamma$-ray TS maps ($8\degr\times 8\degr$ scale with 0.2$\degr$ per pixel) based on Fermi-LAT data between MJD 58072 and MJD 58252. Panel~\ref{tsmaps} corresponds to an analysis source model without 4FGL J0433.0+0522, while in the residual TS map (i.e., panel~\ref{tsmapr}), the target is included. The green X-shaped symbol and ellipse represent optimized position and positional uncertainties of the neutrino, respectively. The red and yellow crosses are the optimized $\gamma$-ray locations of 4FGL J0433.0+0522 and 4FGL J0426.5+0517, respectively, together with the corresponding 95\% C.L. $\gamma$-ray localization error radii (green circles). The white crosses represent other $\gamma$-ray sources in the 4FGL-DR4 catalog. 
}
\end{figure}

Since detection of the neutrino event IC-180213A is highly time-sensitive, we carried out a specific analysis on half-year Fermi-LAT data, centered at the arrival time of the neutrino (i.e., from MJD 58072 to 58252). Such a time length is consistent with the 158-day time window found in the neutrino flare towards to TXS 0506+056 in year 2014-2015 \citep{2018Sci...361..147I}. According to the 4FGL-DR4 \citep{2023arXiv230712546B}, there are two $\gamma$-ray sources likely co-spatial with the neutrino. Besides 3C 120 (i.e., 4FGL J0433.0+0522), another source falls into the localization uncertainty region of the neutrino, 4FGL J0426.5+0517. The analysis on this specific epoch suggests that 4FGL J0426.5+0517 are undetectable for Fermi-LAT then, with given TS values of $<$ 5 (i.e., 1.7$\sigma$). On the other hand, 4FGL J0433.0+0522 is revealed as a significant $\gamma$-ray source, of which the TS value is estimated as 127 (i.e., 11$\sigma$), see Figure \ref{tsmaps}. Its photon flux in this period is obtained as (8.1 $\pm$ 1.2) $\times$ 10$ ^{-8}$ ph cm$^{-2}$ $\rm s^{-1}$. Furthermore, a TS map with 4FGL J0433.0+0522 included in the analysis model was extracted, to check whether sources not in 4FGL-DR4 emerge. As shown in Figure \ref{tsmapr}, no such sources are found. Therefore, when the neutrino arriving, 3C 120 is the unique co-spatial $\gamma$-ray source.

\begin{figure}[ht!]
\centering
\includegraphics[width=\columnwidth]{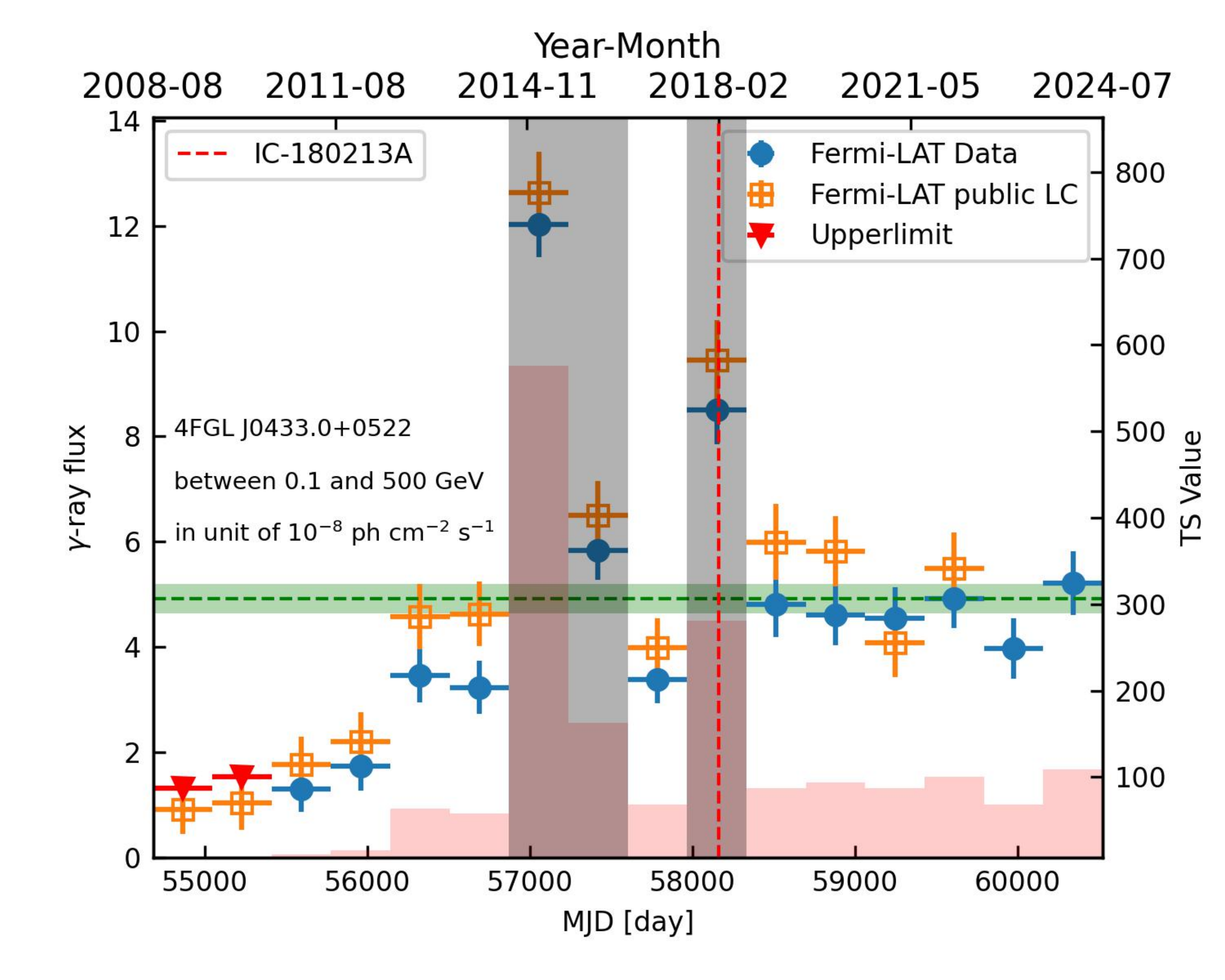}
\caption{
One-year time bin $\gamma$-ray light curve of 4FGL J0433.0+0522. The blue circles and red triangles correspond to flux estimations and upper limits, with TS values indicated by red bars. The orange hollow squares represent data from the Fermi-LAT public Light Curve. The gray shaded region shows the epoch of high-state $\gamma$-ray flux by Bayesian block analyses, and the green shaded region marks the 16-year averaged $\gamma$-ray flux. The red vertical dashed line marks the arrival time of IC-180213A.
}
\label{gylc}
\end{figure}

The temporal behavior of 4FGL J0433.0+0522 is then investigated. Firstly, one-year time-bin $\gamma$-ray light curve was extracted. We are aware of publicly available one-year time bin light curves from the Fermi-LAT collaboration\footnote{\label{LAT-archive}https://fermi.gsfc.nasa.gov/ssc/data/access/lat/14yr\_catalog/4FGL-DR4\_LcPlots\_v32.tgz}, which is based on Summed-Likelihood analyses \citep{2020ApJS..247...33A}. Our results from the Unbinned-likelihood analysis are consistent with that of the public light curve, see Figure~\ref{gylc}. At the beginning of the Fermi-LAT operation, 3C 120 was at a low flux state. Around MJD 56000, its flux started to rise. Two distinct $\gamma$-ray flares were followed, after the flares, it maintained at a median value flux level. The variability is proved to be significant ($>$ $5\sigma$) by adopting the ``variability index" test \citep{2012ApJS..199...31N}. Temporal behaviors of the bright nearby background sources were also investigated. No similar behaviors to those of the target are observed, and hence detections of the $\gamma$-ray flares is likely intrinsic rather than being artificial caused by the backgrounds. Moreover, the flaring epochs are identified by the Bayesian block approach with a false alarm probability of 0.05 \citep{2013ApJ...764..167S}. As shown in Figure~\ref{gylc}, incoming time of IC-180213A (i.e., MJD 58162) coincides with a $\gamma$-ray flare of 4FGL J0433.0+0522 with time range between MJD 57968 and 58333.

\begin{figure}
    \centering
    \includegraphics[width=\columnwidth]{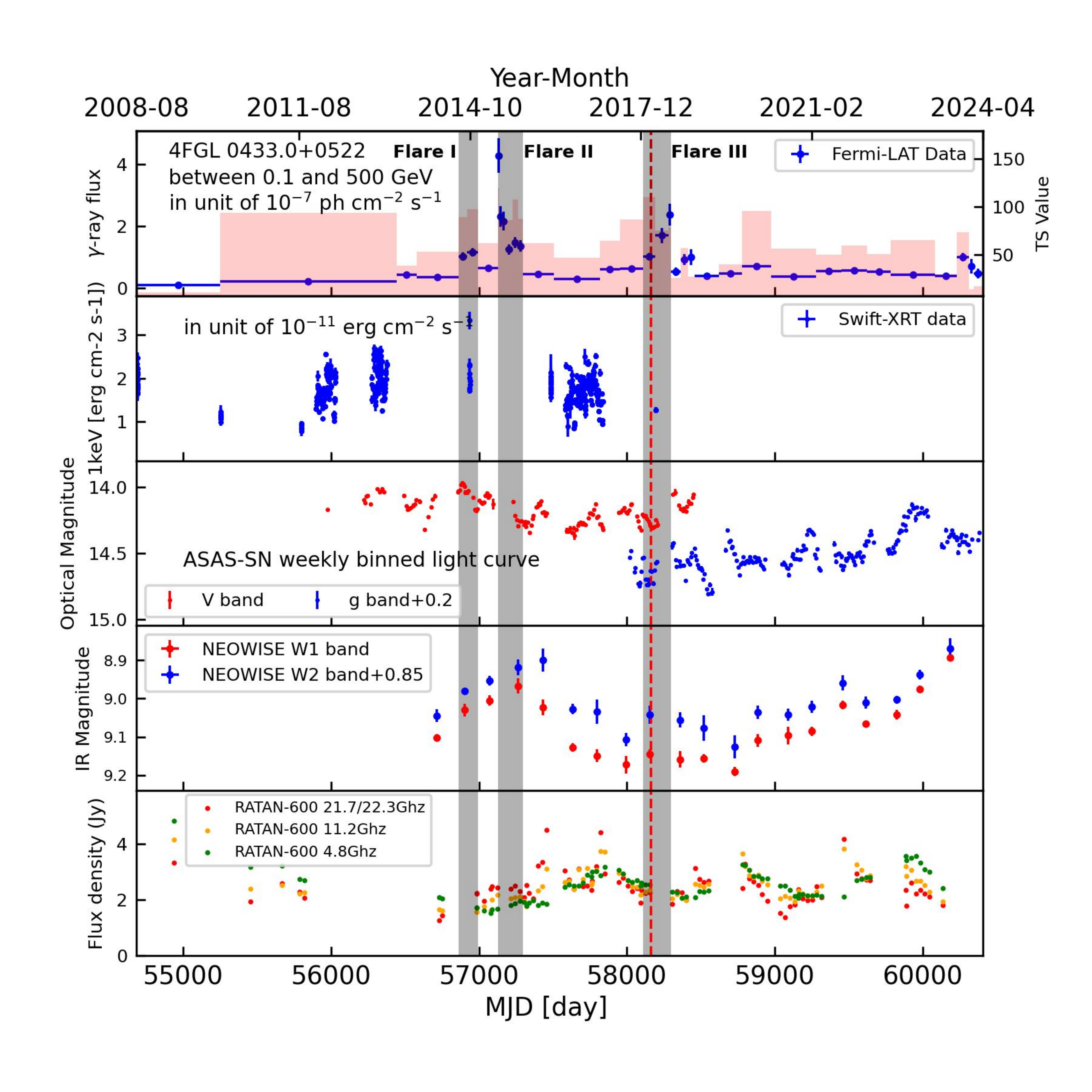}
    \caption{Multi-wavelength light curves of 3C 120. The panels from upper to the bottom are the adaptive-binning Fermi-LAT $\gamma$-ray light curve, the 1 keV light curve from Swift-XRT, the ASAS-SN weekly binned light curves, the NEOWISE light curves, as well as the multi-frequency radio light curves from RATAN-600, respectively. The three shaded gray regions correspond to the time periods of $\gamma$-ray flares, while the red dashed vertical line represents the neutrino arrival time.}
    \label{mlc}
\end{figure}

A constant uncertainty light curve was also extracted, by employing an adaptive-binning method that a default value of target relative flux uncertainty of 20\% was set \citep{2012A&A...544A...6L}. The $\gamma$-ray variation is confirmed to be significant ($>$ $5\sigma$) through the ``variability index" test \citep{2012ApJS..199...31N}. At this case, the comparison between different flux states in long timescale is well characterized, as well as the shape of the $\gamma$-ray flares. Based on the Bayesian Block method \citep{2013ApJ...764..167S}, three significant $\gamma$-ray flares were identified with the false alarm probability of 0.05. These flares, labeled as Flare I, II, and III, are highlighted in Figure~\ref{mlc}.  The most intense flare is the flare II peaking at MJD 57133. Its maximum flux, (4.3 $\pm$ 0.6) $\times$ 10$ ^{-7}$ ph cm$^{-2}$ $\rm s^{-1}$, is roughly 40-fold of the flux level at the initial phase of Fermi-LAT observation around MJD 55000,  (1.0 $\pm$ 0.4) $\times$ 10$ ^{-8}$ ph cm$^{-2}$ $\rm s^{-1}$. The flare III is also distinct, ranging from MJD 58109 to 58298. The peaking flux (at MJD 58290), is (2.4 $\pm$ 0.4) $\times$ 10$ ^{-7}$ ph cm$^{-2}$ $\rm s^{-1}$, which suggests a roughly 20-fold $\gamma$-ray brightening. Note that arrival time of IC-180213A is at the ascent phase of flare III, consistent with the findings from the yearly bin light curve.  By comparison, the flare I appears to be relatively mild, with flux level up to $\simeq 10^{-7}$ ph cm$^{-2}$ $\rm s^{-1}$. In fact, the $\gamma$-ray temporal properties of 3C 120 have been extensively studied. The detections of flare I and II are consistent with the results found in a 30-day time bin light curve \citep{2016MNRAS.458.2360J}. Moreover, adaptive-binning light curves of 3C 120 have been also extracted \citep{2017A&A...608A..37Z,2022MNRAS.510..469K}, from which the variation trends are similar to that shown in this study. Especially, a $\gamma$-ray flux increase is also revealed at the arrival time of IC-180213A \citep{2022MNRAS.510..469K}.

Furthermore, individual analyses on these three flares were performed. The corresponding results are summarized in Table \ref{gflares}. No significant differences of the spectral indexes between these flares are found. Since the spectral index of 4FGL J0433.0+0522 listed by the 4FGL-DR4 is 2.79 $\pm$ 0.03 \citep{2023arXiv230712546B}, no clear spectral hardness is indicated for the flares. Localization analyses suggest that during the flare epochs 3C 120 is co-spatial with the $\gamma$-ray source. Unlike blazars, detections of fast variability from MAGNs are not preferred due to the mild relativistic effect. However, occasionally, a few radio galaxies (e.g., IC 310, \citealt{2014Sci...346.1080A}; NGC 1275, \citealt{2017ApJ...848..111B}) are notable for possessing such extreme observational phenomena. For 3C 120, interestingly, detections of intraday $\gamma$-ray variability in year 2014 - 2015 (i.e., flare I and flare II in our study) have been claimed \citep{2016MNRAS.458.2360J}. It is confirmed in the following study \citep{2017A&A...608A..37Z}. Motivated by these findings, we also extracted an one-day time bin $\gamma$-ray light curve focusing on the epoch of flare III to search for potential fast variability, see Figure~\ref{zoomed-Gamma}. A 3-fold flux rise within one day, from \( (2.1 \pm 0.7) \times 10^{-7} \, \mathrm{ph} \, \mathrm{cm}^{-2} \, \mathrm{s}^{-1} \) at MJD 58191 to \( (6.5 \pm 1.0) \times 10^{-7} \, \mathrm{ph} \, \mathrm{cm}^{-2} \, \mathrm{s}^{-1} \) at MJD 58192, is observed. The TS values of the two detections are 12 (3.4$\sigma$) and 98 (10$\sigma$), respectively. Note that the flux level at MJD 58192 is approximately 13 times of the 16-year averaged flux. Based on these detections, the corresponding doubling timescale in the AGN frame is given as, $\tau_{doub,AGN}=\Delta t\times ln2/ln(F_{1}/F_{2})/(1+z) \lesssim$ 15 hours.

\subsection{Swift-XRT Data}
\label{subsec:2-2}
3C 120 is a well studied bright X-ray source \citep[e.g.,][]{2002Natur.417..625M}. In particular, there are more than 200 visits from the X-ray Telescope aboard Neil Gehrels Swift Observatory \citep{2004ApJ...611.1005G,2005SSRv..120..165B}. Interestingly, one Swift observation on MJD 58196 (ObsID: 00037594050, exposure time of $\simeq$ 1~ks), about one month after the arrival of the neutrino, was performed. The XRT photon counting mode data of this observation were analyzed with the FTOOLS software version 6.33.2. After the xrtpipeline event cleaning by the standard quality cuts, spectra from the target as well as the background were extracted by xselect task. To deal with the pile-up effect, an annular region with an inner radius of 5 pixels and an outer one of 20 pixels was adopted for the source, while a circle with a radius of 60 pixels in a blank area was set for the background. Then, generation of the ancillary response files by the response matrix files taken from the calibration database was followed. The spectrum was re-binned that each bin has at least 20 photons. Channels with energy below 0.5 keV were excluded and the absorption column density was set as the Galactic value (i.e., $\rm 10^{21}~cm^{-2}$). Fitting the spectrum by xspec suggests an unabsorbed 0.5 - 10.0 keV flux of $6.94^{+0.85}_{-0.81}~\times$ 10$ ^{-11}$ erg cm$^{-2}$ $\rm s^{-1}$ ($\chi^{2}$/d.o.f., 19.2/16), as well as a hard x-ray spectrum, $\Gamma_{x} = 1.52 \pm 0.15$.  

The long-term X-ray light curve data were derived from the results from analyses by swift\_xrtproc script \citep{2021MNRAS.507.5690G}\footnote{https://openuniverse.asi.it/blazars/swift/}, in which standard data reduction procedures, corrections for pile-up effect as well as spectral fitting analyses have been employed. Our results for the XRT observation on MJD 58196 are consistent with that given by the swift\_xrtproc script. As shown in Figure \ref{mlc}, at that time, the X-ray flux level is moderate.

\subsection{ASAS-SN Data}
\label{subsec:2-3}
\begin{figure*}[ht!]
\centering
\begin{subfigure}{0.49\textwidth}
    \includegraphics[width=\textwidth]{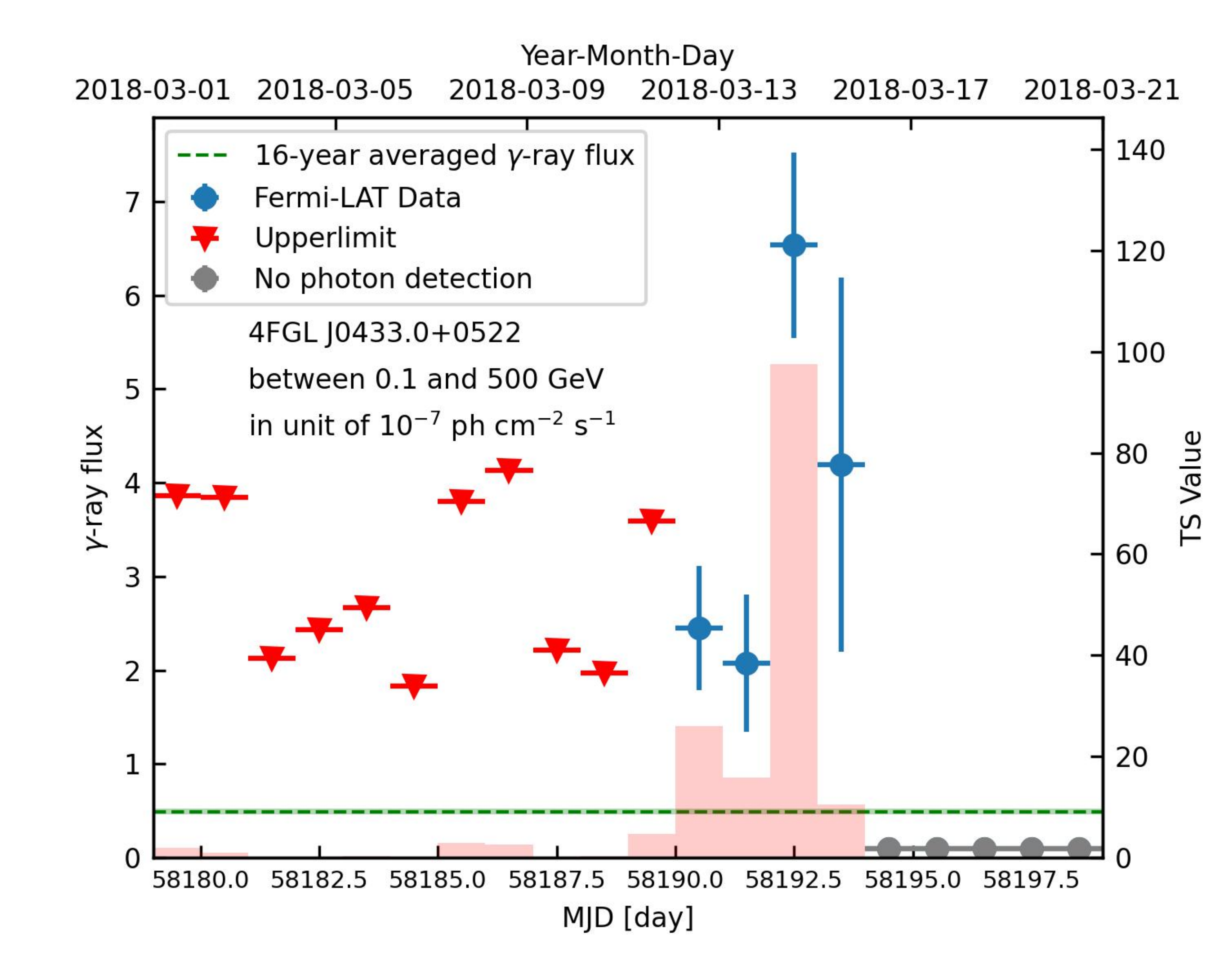}
    \caption{}
    \label{zoomed-Gamma}
\end{subfigure}
\begin{subfigure}{0.49\textwidth}
    \includegraphics[width=\textwidth]{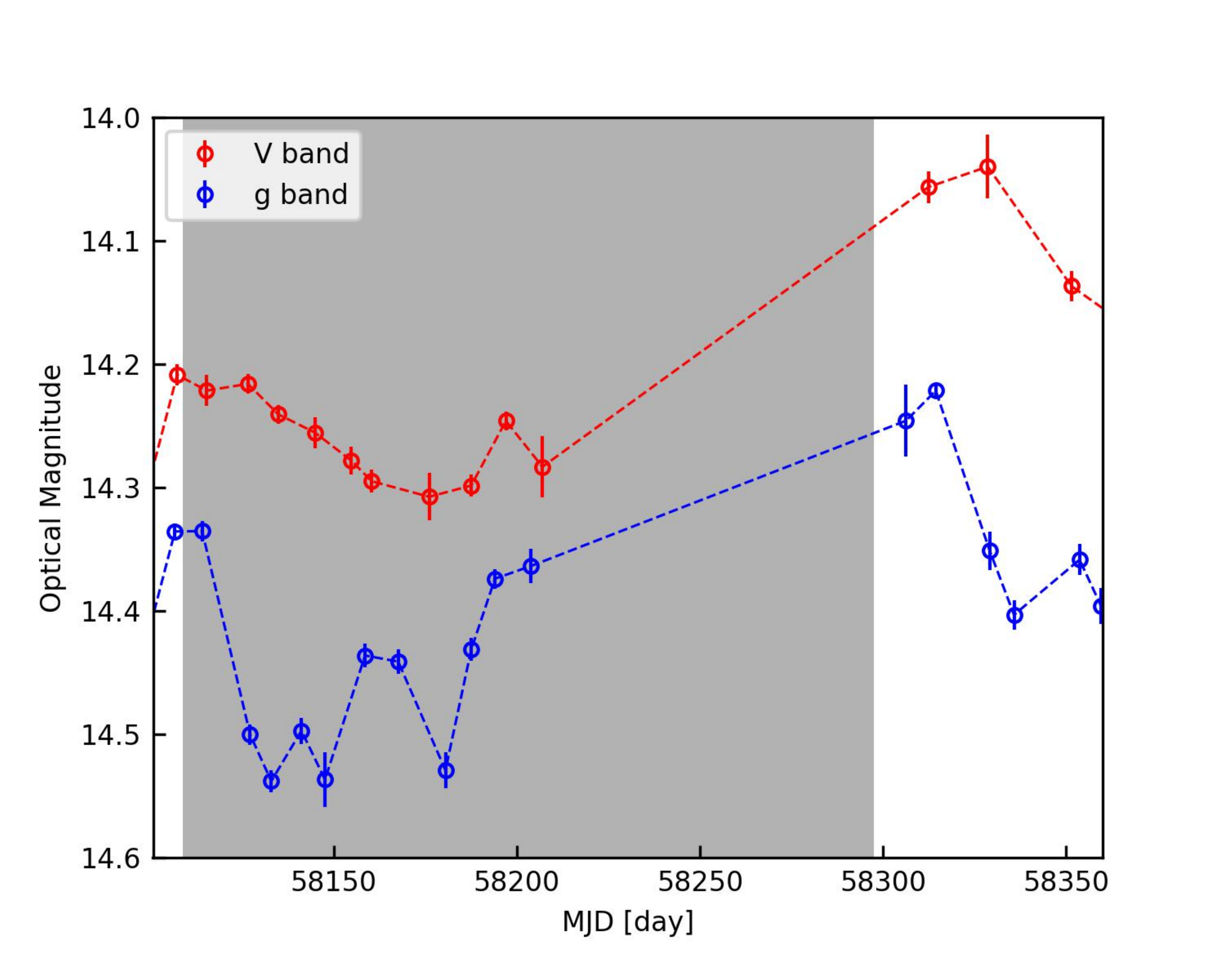}
    \caption{}
    \label{zoomed-Optical}
\end{subfigure}
\caption{The zoomed-in $\gamma$-ray and optical light curves. Left panel: Fermi-LAT one-day time bin $\gamma$-ray light curve during the period from MJD 58179 to 58199. The gray data points indicate time bins with no Fermi-LAT exposures. The 16-year averaged $\gamma$-ray flux is also plotted in the green horizontal line, as well as its uncertainty. Right panel: ASAS-SN light curves in the V and g bands at a time range between MJD 58100 and 58360.
}
\end{figure*}

Optical light curve data were obtained by the All-Sky Automated Survey for Supernovae (ASAS-SN, \citealt{2014ApJ...788...48S,2017PASP..129j4502K,2023arXiv230403791H})\footnote{http://asas-sn.ifa.hawaii.edu/skypatrol/}. V-band and $g$-band magnitudes, extracted by photometry and calibrated by AAVSO Photometric All-Sky Survey \citep{2012JAVSO..40..430H}, from different cameras were adopted. Only frames flagged with quality of ``G" were selected. The exposures were binned into weekly time bin light curves to exhibit the long-term flux temporal behavior. As shown in Figure~\ref{mlc}, the optical emissions of 3C 120 are highly active, with variability amplitudes $ \gtrsim$ 0.5 mag, which is consistent with the results in literature \citep[e.g.,][]{2010ApJ...715..355L}. There are simultaneous ASAS-SN observations, V-band at MJD 58161 and $g$-band at 58160, during the arrival of the neutrino (i.e., MJD 58162). A zoomed-in view of the optical light curves from MJD 58100 to MJD 58360 is shown as Figure~\ref{zoomed-Optical}. A $\chi^{2}$-test that its null hypothesis is a constant flux level, was adopted to investigate the significance of the variability. By varying the constant flux level, minimum $\chi^{2}$ values of the V-band ($\chi^{2}$/d.o.f., 453/13) and the g-band ($\chi^{2}$/d.o.f., 1795/17) light curves are obtained, hence the variability is suggested to be significant (> 5$\sigma$). In particular, a brightening in V-band of 0.27 mag, from 14.31 $\pm$ 0.02 mag at MJD 58176 to 14.04 $\pm$ 0.03 mag at MJD 58328, is detected. Meanwhile, a similar behavior that the $g$-band flux rises from 14.53 $\pm$ 0.01 mag at MJD 58180 to 14.22 $\pm$ 0.01 mag at MJD 58314 is also found. Since the flux increases are markedly larger than the measurement uncertainties, $\Delta mag/\sqrt{(magerr^{2}_{1}+magerr^{2}_{2})} \gtrsim 7 $, the brightenings are likely intrinsic rather than due to the fluctuation. However, there are no simultaneous ASAS-SN observations when the $\gamma$-ray flux reaches to the maximum value (i.e. MJD 58290). The temporal relationship between these two bands cannot be well constrained.

\begin{table*}[ht!]
\centering
\caption{The parameter values obtained from Fermi-LAT data analysis during \textbf{Flare I}, \textbf{Flare II}, and \textbf{Flare III} as shown in Figure~\ref{mlc}.}
\label{gflares}
\begin{tabular}{lccc}
\hline\hline
\textbf{Results} & \textbf{Flare I} & \textbf{Flare II} & \textbf{Flare III} \\ \hline
Period (MJD) & 56861 – 56993 & 57129 – 57298 & 58109 – 58298 \\
TS & 179 & 577 & 200 \\
Averaged Flux (10$^{-7}$ ph cm$^{-2}$ s${-1}$) & 1.09 $\pm$ 0.12 & 1.81 $\pm$ 0.11 & 1.21 $\pm$ 0.13 \\
Peaking Flux (10$^{-7}$ ph cm$^{-2}$ s${-1}$) & 1.16 $\pm$ 0.14 & 4.27 $\pm$ 0.56 & 2.37 $\pm$ 0.35 \\
Spectral Index & 2.67 $\pm$ 0.09 & 2.68 $\pm$ 0.06 & 2.68 $\pm$ 0.09 \\
Highest Energy Photon (association probability) & 2 GeV (0.91)  & 5 GeV (0.91)  & 3 GeV (0.92)  \\ 
\hline\hline
\end{tabular}

\end{table*}

\subsection{WISE data}
\label{subsec:2-4}
Single-exposure photometric data in the W1 and W2 bands (centered at 3.4 and 4.6 $\mu$m in the rest frame) from the Wide-field Infrared Survey Explorer (WISE; \citealt{2010AJ....140.1868W}) and the NEOWISE Reactivation mission \citep{2020DPS....5220801M,neowise} were collected. Data with poor image quality (“ qi\_fact” $<$ 1), close proximity to the South Atlantic anomaly (“ SAA” $<$ 5), as well as those flagged for moon masking (“ moon mask” $=$ 1) were filtered out \citep{2021ApJS..252...32J}. We grouped the data for each observational epoch (see Figure~\ref{mlc}), separated by approximately half a year, using the weighted mean value. The corresponding uncertainty was determined by propagation of the measurement errors. An investigation based on a $\chi^{2}$-test statistic analysis (for more details, see \citealt{2019ApJ...879L...9L}) suggests significant long-term variability ($>$ 5$\sigma$). The variability amplitude is up to $\simeq$ 0.3 mag. One flux maximum is distinct at MJD 57264. A continuous flux increase is exhibited in recent years and the flux is at a high flux level right now. There are WISE observations on MJD 58154, about 8 days before the detection of the neutrino. The infrared fluxes at that time are at a relatively low flux state.

\subsection{RATAN-600 data}
\label{subsec:2-5}
Additional radio data were collected from observations from RATAN-600, a 600-meter circular multi-element antenna that is capable of measuring broad-band spectra across the 1–22 GHz range simultaneously \citep{1993IAPM...35....7P}, since 3C 120 has been included in the RATAN-600 multi-frequency monitoring list for jetted AGNs\footnote{http://www.sao.ru/blcat/} \citep{2014A&A...572A..59M,2022AstBu..77..361S}. As shown in Figure~\ref{mlc}, the centimeter radio fluxes are variable, especially at 21.7/22.3~GHz in which the flux density at MJD 57457 (4.5 $\pm$ 0.6 Jy) is three times of that at MJD 56731 (1.3 $\pm$ 0.2 Jy). RATAN-600 observations at 21.7/22.3, 11.2 and 4.8~GHz have been performed on MJD 58156, about 6 days before the detection of IC-180213A. The measured flux densities then are relatively low.

\subsection{Implications of multi-messenger observations}
\label{subsec:2-6}

The broadband temporal behaviors, shown in Figure~\ref{mlc}, strongly suggest that the activities of the relativistic jet are distinct from radio wavelengths to the high energy $\gamma$-ray regime, although as a MAGN the Doppler effect of 3C 120 is supposed to be relatively low. In particular, the variability in the GeV $\gamma$-ray domain is more violent than that in optical/infrared wavelengths, and in the latter case, contributions from the accretion disk as well as the host galaxy are likely also significant. Since three flares have been identified in the $\gamma$-ray light curve, the temporal behaviors then at other radiation windows are examined. For the flare III, in its initial phase of the flux rise when the neutrino event IC-180213A is detected, brightening of the optical fluxes in the both $g$-band and V-band are followed. However, no observations in X-ray, and infrared around the peaking time of flare III are available. Although the variability amplitude of flare I is moderate, at the same time a strong rapid X-ray flare appears.

The multi-wavelengths electromagnetic observations play an important role on investigation of origin of the neutrino event IC-180213A. Aforementioned evidences include that in a half-year epoch centered at the arrival time of the neutrino 3C 120 is the unique known co-spatial $\gamma$-ray source. In addition, the incoming of IC-180213A temporally coincides with the second strongest $\gamma$-ray flare among the entire 16-yr observations, of which 20-fold flux increase is found. Accompanying optical flux increases are also detected. All these observational facts suggest a physical link between IC-180213A and 3C 120.

Further Monte Carlo simulations were conducted to estimate the chance probability corresponding to the multi-messenger detections. The central position of IC-180213A was randomized across the sky, by fixing the size of localization error area \citep{2023ApJS..269...25A}. Since the detection sensitivity of IceCube is highly dependent on the sky declination, only the R.A. value was left free. Since the neutrino (b < $-26 ^{\circ}$) is away from the Galactic plane, it is unlikely from the Galactic sources \citep{2024ApJ...969..161G,2024ApJ...975L..35F}. Therefore, during the simulation, the Galactic plane region at this sky declination was excluded, about 10\% of the total considered area. Known $\gamma$-ray sources, showing an outburst when IC-180213A came, were searched in the simulation. Based on the public one-year time bin Fermi-LAT light curves,  only sources with significant variability (p > 99\%, \citealt{2020ApJS..247...33A}) were selected. An additional criterion is that they should be significantly detected (TS > 25) by Fermi-LAT when the neutrino came (i. e., the tenth time bin of the Fermi-LAT light curves). The selections yield 21 sources and extraction of their adaptive-binning light curves were performed. Considering that a three-time flux increase compared to the long timescale averaged flux level is usually adopted to define a major flare \citep[e.g.,][]{2011ApJ...733L..26A}, three sources, TXS 0506+056, GB6 J0922+0433 and TXS 1421+048, were picked out. The chance probability is calculated as $p = \frac{M+1}{N+1}$, where $N$ represents the total number of times of simulations (i.e., $10^{4}$) and $M$ denotes the number of cases that temporally coincident sources (but not co-spatial) fall within the simulated uncertainty region. Under this circumstance, p $\simeq$ 0.04 is yielded, hence 3C 120 is suggested as a possible neutrino emitter. However, note that the calculated significance is a posteriori and it should be considered informative rather than definitive. Moreover, in order to check whether the exclusion of the low galactic latitude sources could influence the probability, their $\gamma$-ray temporal properties have been investigated. There are in total 37 4FGL-DR4 sources in this direction, in which only three ones exhibit significant long-term variability and none of them possess a clear flux rise when the neutrino arrived. Therefore, the estimated probability holds no matter whether the galactic plane region is excluded.

Since 3C 120 is one of a few nearby bright $\gamma$-ray radio galaxies, investigations of the origin of its high-energy emission has been carried out. A time lag of dozens of days between the $\gamma$-ray and 43~GHz radio light curves suggests that the the $\gamma$-ray emission region is at sub-parsecs scales from the SMBH and synchrotron self-Compton processes (SSC) provide a well description of the $\gamma$-ray emissions \citep[e.g.,][]{2015A&A...574A..88S,2015ApJ...799L..18T}. On the other hand, observed fast intraday $\gamma$-ray variability indicates a deeply embedded emission region in the flaring epochs ($\sim$ 0.002 pc, \citealt{2016MNRAS.458.2360J}), and external Compton processes cannot be ruled out \citep[e.g.,][]{2017A&A...608A..37Z}. The previous studies are based on multi-wavelength data from year 2012 to 2015, in our work, new information have been given. Taking the 15-hour doubling time in flare III as an example, the radius of the radiation region is constrained as, $r < c  \delta \tau_{doub,AGN} \approx 0.0013$ pc, where a Doppler boosting factor of \( \delta = 2.4 \) is adopted \citep{2005AJ....130.1418J}. Under a conical jet geometry, the distance between the radiation region and the central SMBH is suggested as \( r_{\text{b}}  = \Gamma r \lesssim 0.007\) pc, where \(\Gamma\) is taken to be 5 \citep{2005AJ....130.1418J}. Meanwhile, the time lag measured in the reverberation mapping observations between the emission lines and the continuum radiations is given as $\simeq$ 28 light days \citep{2014A&A...568A..36P}, indicating that the broad line region (BLR) is at a distance of $\sim$ 0.02 pc away from the central SMBH. Therefore, in the flare III, the jet radiation region is likely within the BLR. It is also worth noting that  the most energetic photons of 3C 120 at energies of a few GeVs, see Table \ref{gflares}. The possible absorption by UV photons from the BLR could explain the absence of $\gamma$ rays at energy of tens of GeVs \citep{2010ApJ...717L.118P}.

\begin{figure}[ht!]
\centering
\includegraphics[width=\columnwidth]{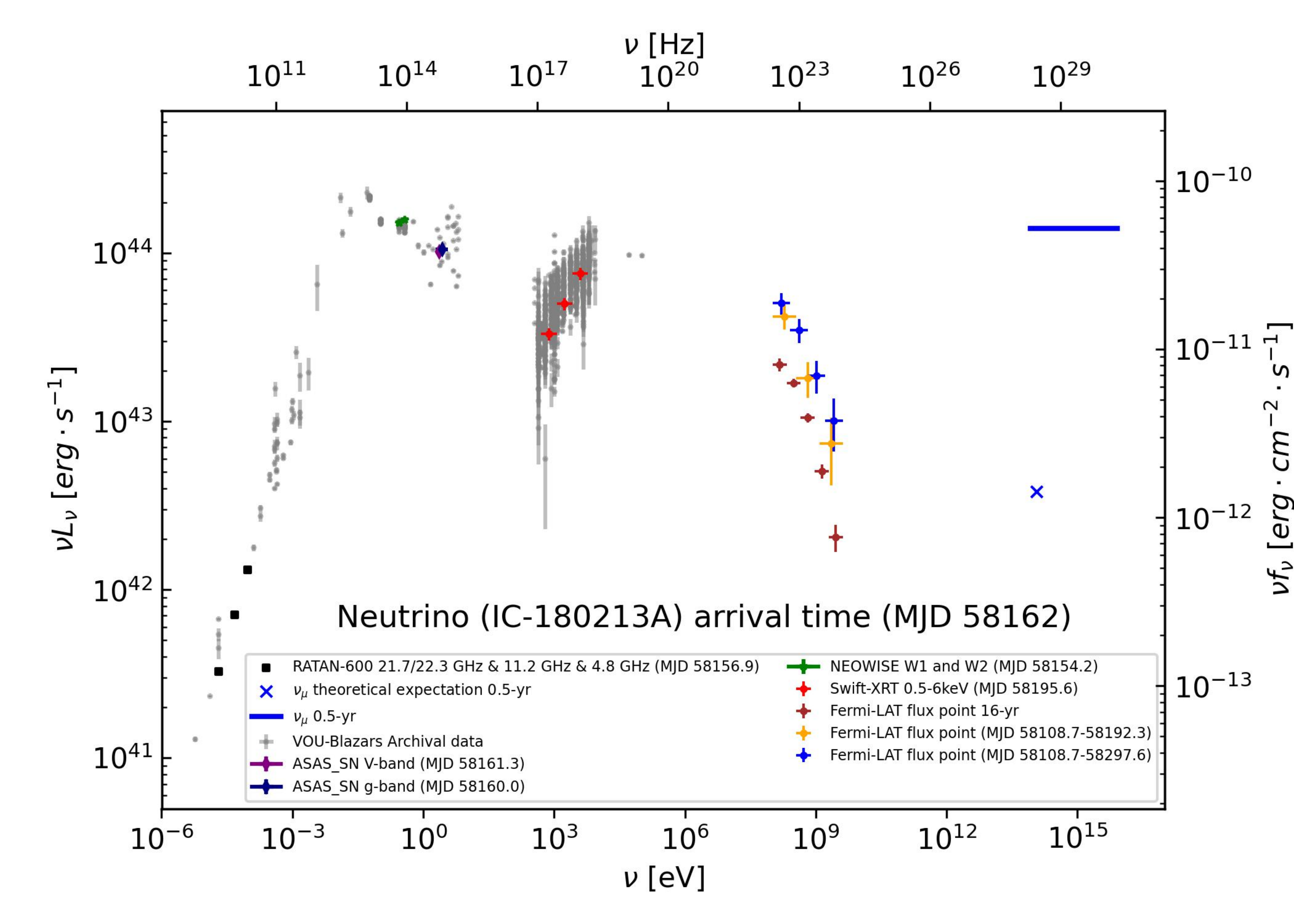}
\caption{The broadband SED of 3C 120 quasi-simultaneous data are displayed in color. The horizon line indicates the energy-averaged neutrino energy flux over a six-month duration. In contrast, the X-shaped symbols denote theoretical expectations of neutrinos. Non-simultaneous archival data are in gray.}
\label{sed}
\end{figure}

The spectral energy distribution (SED) of 3C 120 is drawn in Figure~\ref{sed}, where a typical two-bump shape is exhibited. Thanks to multi-wavelength time-domain observations, (quasi)simultaneous detections (i.e., within one week) in radio, infrared, and optical bands are available, coinciding with the arrival of the neutrino event. Additionally, archival non-simultaneous data \citep{2020A&C....3000350C} are included as background for comparison. The (quasi)simultaneous SED reveals that the $\gamma$-ray emission is elevated compared to the 16-year averaged measurement, while flux levels in other wavebands appear moderate. The synchrotron peak frequency of 3C 120 is constrained to $\nu_{s} \simeq 10^{13}~\text{Hz}$, which suggests it as a low-synchrotron-peaked source \citep{2010ApJ...716...30A}.

The frequency of the synchrotron emission can be derived as, $\rm \nu_{syn,jet} = \nu_{syn,obs} (1+z)/\delta \simeq \gamma^{2}eB'/2\pi m_{e}c$, where $\gamma$ is the energy of emitting electrons, $\rm \nu_{syn,jet}$ is in jet co-moving frame and $\rm \nu_{syn,obs}$ is in observational frame \citep{2013LNP...873.....G}. Assuming that the emitting electrons follow a broken power-law distribution, the electrons at the spectral breaking energy (i.e., $\gamma_{b}$) are expected to significantly contribute the emission at the synchrotron peak. Therefore, the magnetic field strength is constrained as $\rm B' = \nu_{syn,obs}(1+0.033)/(3.7\times 10^{6}\gamma_b^2 \delta$) \citep{1998ApJ...509..608T}. By setting $\gamma_{b} = 10^3$ which is consistent with values used in literature \citep[e.g.,][]{2015ApJ...799L..18T,2016MNRAS.458.2360J}, and adopting $\delta = 2.4$, the magnetic field strength is implied as $\sim 1~\text{Gauss}$. Note that in addition to the synchrotron emission the electrons also are responsible to the $\gamma$-ray emissions which are likely vary quickly. The estimated magnetic field strength corresponds to the compact jet blob.

However, given its soft $\gamma$-ray spectrum and the lack of hard X-ray data, the peak frequency of the high energy bump cannot be well constrained. As mentioned above, the radiation mechanisms responsible for the $\gamma$-ray emissions of 3C 120 is still unclear. Moreover, if multi-messenger information are needed to be understood, thorough lepto-hadronic radiation theoretical calculations should be performed, which is beyond the scope of this work. Nevertheless, it is agreed that leptonic processes are mainly responsible for the production of the $\gamma$-ray radiation of 3C 120. The leptonic scenario gives a natural explanation of the quasi-simultaneous flux increase of the optical and $\gamma$-ray bands in year 2018 (for a similar case, see \citealt{2014ApJ...783...83L}). However, if the neutrino is indeed from the 3C 120, a secondary hadronic component in the $\gamma$-ray regime, especially in flare III of our study, is needed to be also considered.

\section{Discussions and summary} \label{sec:3}

According to the detection of IC-180213A, a constraint on the neutrino luminosity of 3C 120 can be given. The number of (anti-)muon neutrinos at a declination $ \delta $ detected by IceCube over a time period $\Delta T$ is expressed as,
\begin{equation}
     {N_{\nu_{\mu}}}  = \Delta{T} \int_{\epsilon_{\nu_{\mu, \text{min}}}}^{\epsilon_{\nu_{\mu, \text{max}}}} {A_{\text{eff}}(\epsilon_{\nu_{\mu}}, \delta) \phi_{\nu_{\mu}} } \, d\epsilon_{\nu_{\mu}}.
\end{equation}
The energy range of the neutrino population is set between \( \epsilon_{\nu_\mu,\text{min}} = 80 \, \text{TeV} \) and \( \epsilon_{\nu_\mu,\text{max}} = 8 \, \text{PeV} \), where detections of 90\% of neutrinos in the $\gamma$-ray follow-Up (GFU) channel are expected \citep{2021JCAP...10..082O}. The neutrino spectrum is characterized by a power-law distribution \( \epsilon^{-\gamma} \), with an index \( \gamma = 2 \). Meanwhile, \( \phi_{\nu_\mu} \) represents the differential muon neutrino flux. A GFU\_Bronze type event (i.e., IC-180213A), the corresponding effective area is approximately \( A_{\text{eff}} \approx 16 \, \text{m}^2 \) \citep{2023ApJS..269...25A}. For the integrated time $\Delta T$, we take a duration of half-year when the $\gamma$-ray emission is at a high flux level, identified by Bayesian Blocks analysis of the adaptive-binning light curve. In principle, a scenario of injections of fresh relativistic particles into the radiation zone is usually used to interpret the activities of AGN jets. The short radiative cooling time of electrons leads to significant electromagnetic radiation outputs, meanwhile, an accompanying rise of neutrino flux by the accelerated protons are also expected  \citep[e.g.,][]{2016ApJ...831...12H,2016APh....80..115P,2018ApJ...865..124M,2020ApJ...902..133K}. Therefore, the integrated muon neutrino energy flux is calculated as approximately \(2.4 \times 10^{-10} \, \text{erg} \, \text{cm}^{-2} \, \text{s}^{-1} \). Considering the proximity of 3C 120, the integrated muon neutrino luminosity is obtained as \( L_{\nu_\mu} \approx 6.4 \times 10^{44} \, \text{erg} \, \text{s}^{-1} \), or alternatively, an average muon neutrino luminosity \( \epsilon_{\nu_\mu} L_{\epsilon_{\nu_\mu}} = \frac{L_{\nu_\mu}}{\ln(8 \, \text{PeV} / 80 \, \text{TeV})} \approx 1.4 \times 10^{44} \, \text{erg} \, \text{s}^{-1} \).

In this work, the photo-pion production (\( p + \gamma \rightarrow p + \pi \)) is taken as the responsible mechanism to the neutrino production. The energy in the cosmic rest frame for IC-180213A is \( \epsilon_{\nu} = \epsilon_{\nu,obs}(1+z) \approx 0.115 \,\text{PeV} \)\footnote{Quantities with subscripts ‘obs’ pertain to the observer’s frame, while those with primes refer to the frame co-moving with the jet. All other ones are in the cosmic rest frame unless stated otherwise.}.  Hence the energy of emitting proton in the co-moving frame is at \( \epsilon'_{p} \approx 20 \epsilon_{\nu}/\delta \approx 1 \, \text{PeV} \). When the meson production is dominated by the $\Delta$-resonance and direct pion production, the energy of the targeting photons in the co-moving frame is suggested as \( \epsilon'_{t} \approx 0.125\,\text{keV} \left( \frac{2.4}{\delta} \right) \left( \frac{0.115\,\text{PeV}}{\epsilon_{\nu}} \right) \) \citep{2018ApJ...865..124M}. If the photons are external of the jet, the corresponding photon energy in the cosmic rest frame is given as \( \epsilon_{t} = \epsilon'_{t}/ \Gamma \sim 25 \,\text{eV}\). Considering that 3C 120 is defined as a BLRG and its accretion disk emission is luminous, $\sim 9\lambda L\lambda[5100\AA] \sim 1.3\times 10^{45}$ erg $\rm s^{-1}$ \citep{2004ApJ...613..682P}, the UV photons from the BLR likely act as the soft photons for the photo-pion processes.  

During the photo-pion interaction, protons transfer 3/8 of their energy to neutrinos, while the rest 5/8 part corresponds to the production of electrons and pionic $\gamma$ rays. Subsequent electromagnetic cascades are initiated, leading a potential contribution on the detected electromagnetic radiations. Note that synchrotron cooling is likely dominant here, due to the Klein-Nishina suppression in the inverse Compton scattering. The connection between neutrino radiation and related $\gamma$-ray emission from the cascade is suggested in \cite{2018ApJ...865..124M}, 

\begin{equation}
    {\epsilon_\nu}{L_{\epsilon_{\nu}}} \approx \frac{6\left(1+Y_{IC}\right)}{5}{\epsilon_\gamma}{L_{\epsilon_{\gamma}}}\vert_{\epsilon_{\text{syn}}^{p\pi}} \approx 1.2 \times 10^{43} \, \text{erg s}^{-1} \left(\frac{{\epsilon_\gamma}{L_{\epsilon_{\gamma}}}\vert_{\epsilon_{\text{syn}}^{p\pi}}}{10^{43}}\right).
\end{equation}

\( Y_{\text{IC}} \) represents the Compton-Y parameter for pairs from the cascades, which is typically \( \leq 1 \) \citep{2018ApJ...865..124M}. If a Doppler factor value of 2.4 as well as a magnetic filed intensity of \(B' = 1~\text{Gauss} \) are set, the synchrotron radiations by the secondary pairs from the cascades peak at, \( \epsilon_{\text{syn,obs}}^{p\pi} \approx 0.3 \, \text{GeV} \left( \frac{B'}{1 \, \text{Gauss}} \right) \left( \frac{\epsilon_p}{2.3 \, \text{PeV}} \right)^2 \left( \frac{2.4}{\delta} \right) \left( \frac{1.033}{1 + z} \right) \). In the most optimistic scenario, the observed $\gamma$-ray emissions at such an energy could be overwhelmed by emissions from the secondary pairs. Or conservatively, emissions from the primary leptons are dominant. Since the typical relative uncertainty for Fermi-LAT estimations is roughly 20\%, and the observed luminosity at 0.3~GeV is about $5 \times 10^{43} \, \text{erg s}^{-1} \) (shown in Figure~\ref{sed}), ${\epsilon_\gamma}{L_{\epsilon_{\gamma}}}\vert_{\epsilon_{\text{syn}}^{p\pi}}$ = $10^{43}$ erg $\rm s^{-1}$ is assumed. Therefore, a (anti-)muon neutrino luminosity of \(4 \times 10^{42} \, \text{erg s}^{-1}\) is expected. Compared to the luminosity of muon neutrinos inferred from the detection of IC-180213A, the probability of observing such a neutrino due to Poisson fluctuations is approximately 0.03.

\begin{figure*}[ht!]
\centering
\begin{subfigure}{0.45\textwidth}
    \centering
    \includegraphics[width=\textwidth]{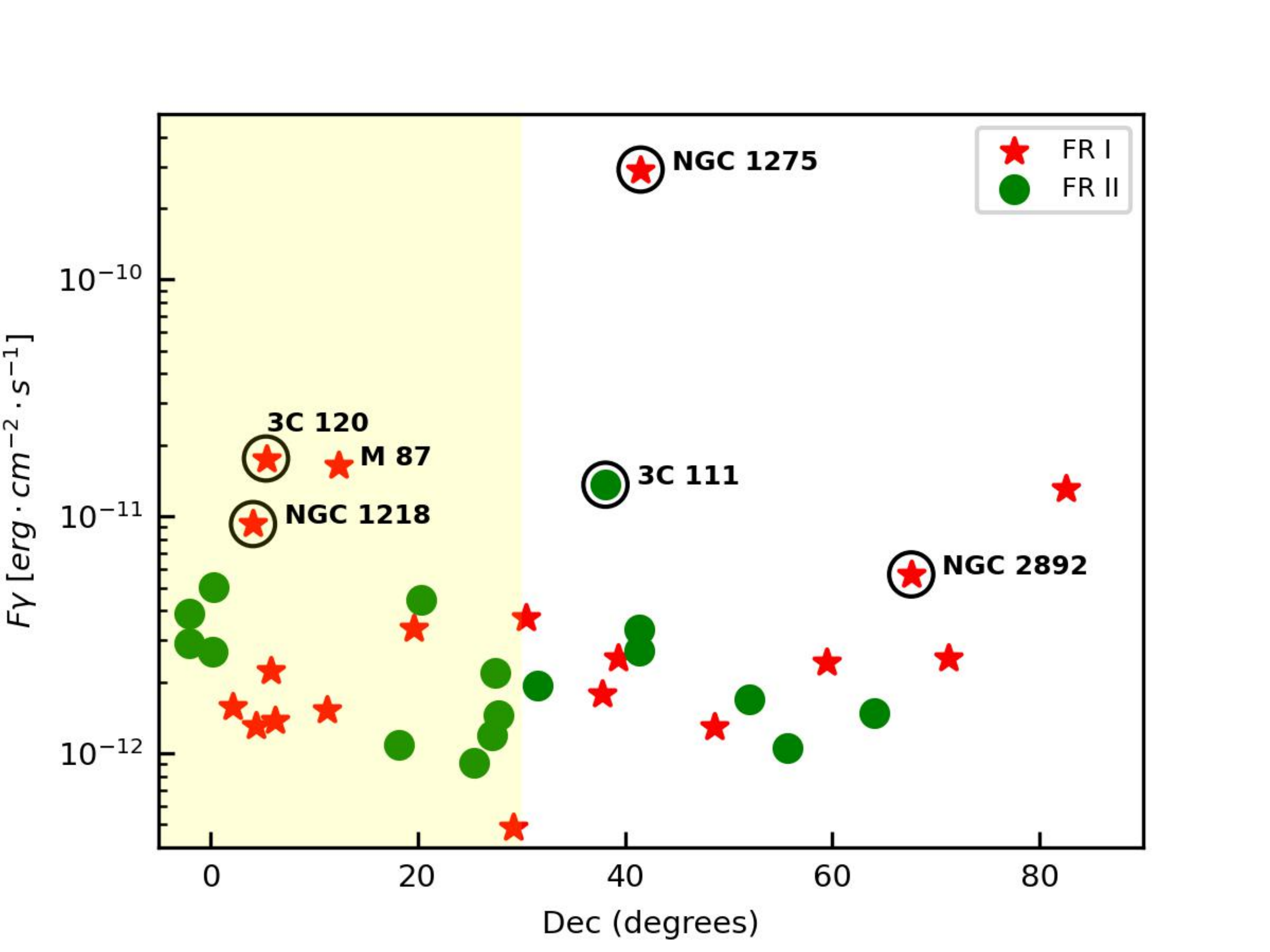}
    \caption{}
    \label{GLR}
\end{subfigure}
\hspace{0.05\textwidth}
\begin{subfigure}{0.45\textwidth}
    \centering
    \includegraphics[width=\textwidth]{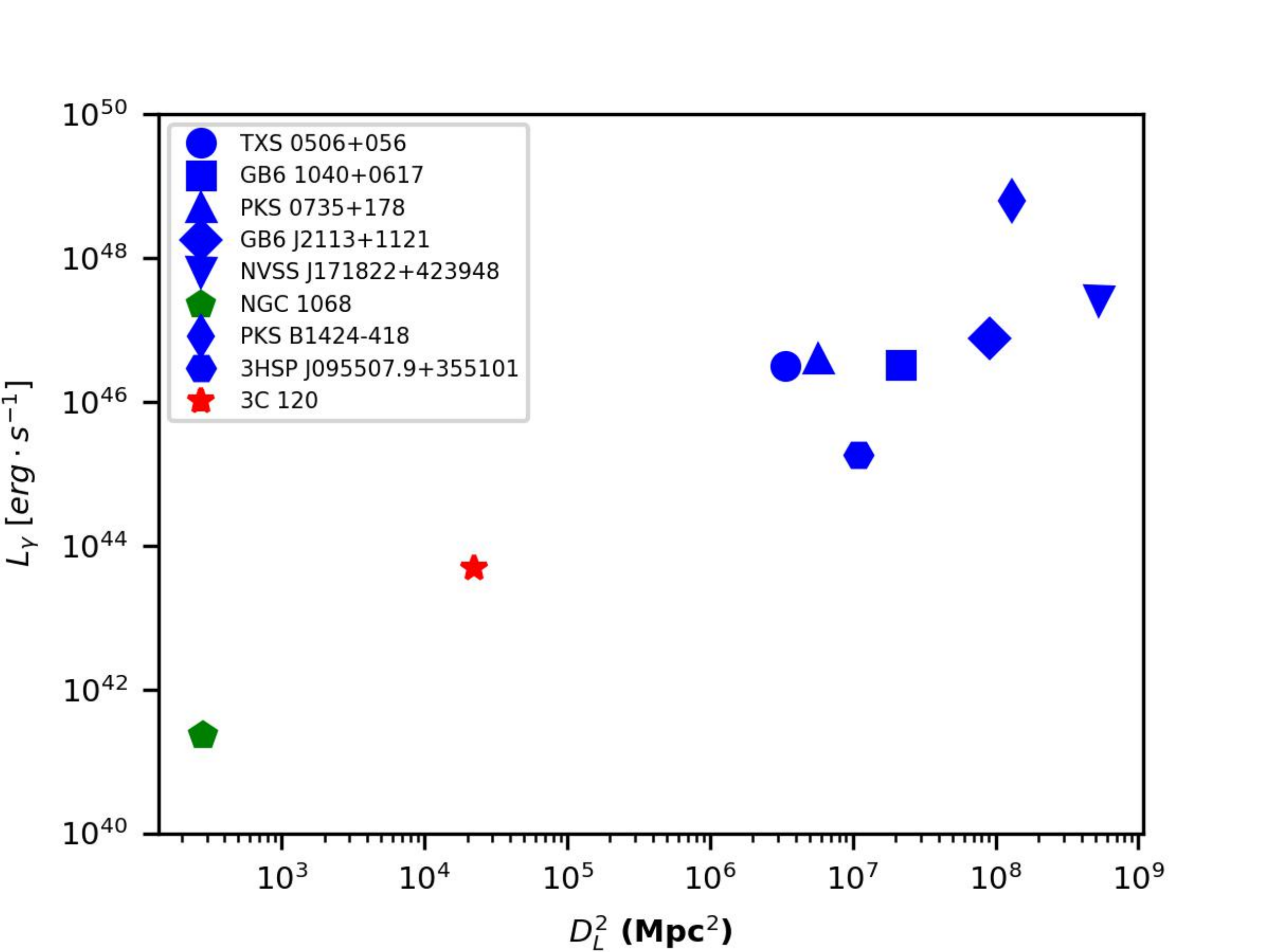}
    \caption{}
    \label{GLN}
\end{subfigure}

\caption{
(a): The comparison of 3C 120 with other GeV $\gamma$-ray radio galaxies. The red asterisks represent FR I radio galaxies, while the green circles denote FR II ones. The sources enclosed by black circles demonstrate significant (> 99\%) long-term GeV $\gamma$-ray variability \citep{2023arXiv230712546B}. Note that M 87 exhibits pronounced variability in the TeV $\gamma$-ray domain. The yellow shaded area represents the optimal declination range for IceCube's effective area. 
(b): The comparison between 3C 120 and the neutrino-emitting candidates. The blue shapes represent blazars, the red shapes correspond to the radio galaxy 3C 120, while the green shapes denote the Seyfert II galaxy NGC 1068. Since other radio quiet Seyfert galaxy candidates are not included in 4FGL-DR4 \citep{2023arXiv230712546B}, they are not plotted here.}
\label{fig:comparison}
\end{figure*}

Moreover, investigations on the kinetic energy of the jet based on the detection of the neutrino have been performed. As mentioned above, an all-flavor neutrino luminosity is described by \( \epsilon_{\nu} L_{\epsilon_{\nu}} = \frac{3}{8} f_{p\pi} \epsilon_{p} L_{\epsilon_{p}} \), where the optical depth for \( p\pi\) processes (\( f_{p\pi} \)) accounts for the efficiency of neutrino production. For 3C 120, BLR photons, which are described as a blackbody distribution with a temperature of 42000 K, are considered as target photons in the photo-pion processes. The optical depth for \( p\pi \) processes is approximated by \( f_{p\pi} \approx \hat{n}_{\text{BLR}} \sigma_{p\pi}^{\text{eff}} r_{\text{b}}  \approx 2.8 \times 10^{-4} \) \citep{2014PhRvD..90b3007M}, where \( \hat{n}_{\text{BLR}} \approx 2 \times 10^8 \, \text{cm}^{-3} \) denotes the photon density at $\epsilon_t$ = 25~eV, \( \sigma_{p\pi}^{\text{eff}} = 0.7 \times 10^{-28} \, \text{cm}^2 \) is the effective cross-section for \( p\pi \) interactions \citep{2016PhRvL.116g1101M}, and \( r_{\text{b}} \) represents the location of the emission region. Corresponding to the flare III, the proton luminosity is constrained as \(\epsilon_p L_{\epsilon_p}  \approx 3.6 \times 10^{46} \, \text{erg s}^{-1} \). Assuming a power-law spectrum for protons (\(\epsilon_p^{-\gamma}\) with \(\gamma = 2\), \(\epsilon_{p,\text{min}} = \delta m_p c^2\), and \(\epsilon_{p,\text{max}} = 10^{18} \, \text{eV}\), where \(m_p c^2\) is the proton rest energy), together with \(\delta = 2.4\), the proton luminosity \(L_p\) is given by \( \epsilon_p L_{\epsilon_p} \ln \left(\frac{10^{18} \, \text{eV}}{\delta m_p c^2}\right) \approx 7.2 \times 10^{47} \, \text{erg s}^{-1} \). In the AGN frame, the jet proton luminosity \(\hat{L}_{p,\text{jet}}\) is approximately \( L_p / (4 \Gamma^{2}/3) \approx 2.2 \times 10^{46} \, \text{erg s}^{-1} \). Such a value is about three times of the Eddington luminosity, \(7 \times 10^{45} \, \text{erg s}^{-1}\), consistent with the results on other neutrino emitting candidates \citep[e.g.,][]{2019MNRAS.483L..12C,2019NatAs...3...88G,2021ApJ...912...54R}.

Alternative external photon scenarios are worthy of discussions. One option is that the emission region is at the vicinity of the SMBH, where UV/X-ray photons from the disk/corona are intense. Although in such an extreme environment,  productions of neutrinos could be efficiently enhanced \citep[e.g.,][]{2020ApJ...891L..33I,2020PhRvL.125a1101M}, the attenuation of GeV/sub-GeV $\gamma$-ray photons is likely severe. Therefore, the temporal coincidence between a rise of the $\gamma$-ray flux and arrival of a neutrino is not anticipated. On the other hand, the emission region could be distant from the central engine, at sub-parsec away, where the infrared emissions from the dust torus are the external photons. Under this circumstance, the photon number density is orders of magnitudes lower than that from emissions of the BLR, and hence the production of neutrinos is suppressed. Moreover, considering a typical energy of photons from the hot dust, roughly 0.5 eV, the energy of the corresponding IceCube neutrino would be at 5~PeV, much higher than that of IC IC-180213A, 0.111 PeV. Conclusively, the UV photons from the BLR appear to be the suitable soft photons for the photo-pion processes.

So far there are more than 60 MAGNs detected by Fermi-LAT, that majority of them are radio galaxies \citep{2023arXiv230712546B}. As shown in Figure~\ref{GLR}, 3C 120 is one of the brightest radio galaxies. Meanwhile, it is one of the few ones in its kind possessing significant GeV $\gamma$-ray variability. Furthermore, 3C 120 is at the horizon direction of the IceCube, that is ideal for the IceCube detection. For a neutrino at energy of 0.111 PeV like IC-180213A,  the GFU Bronze+Gold effective area at a declination value of several degrees (i.e., 0$\degr$ < $\delta$ < 30$\degr$), is orders of magnitudes larger than that at the deep southern hemisphere (i.e., $\delta$ < - 5$\degr$), and slightly larger than that in other directions (i.e., $\delta$ > 30$\degr$ as well as -5$\degr$ < $\delta$ < 0$\degr$). Therefore, it is not surprising that 3C 120 is the first radio galaxy proposed as a neutrino emitter. Besides 3C 120, there are a few other radio galaxies are worth noting for future neutrino detections. The first one is NGC 1275, the most powerful and active radio galaxy in the GeV domain \citep{2023arXiv230712546B}. However, since the declination angle of NGC 1275 is +41.3$ ^{\circ}$, its IceCube effective area could be significantly reduced compared with 3C 120 if the energy of the arrival neutrino reaches to $\simeq$ 1~PeV \citep{2023ApJS..269...25A}. Other sources include M 87 and NGC 1218. Strong TeV variations have been detected in M 87 \citep{2008ICRC....3..937B}, while the GeV $\gamma$-ray behavior and the sky declination of the latter are similar to 3C 120.

In the nearby universe (i.e., z $<$ 0.05), in addition to NGC 1068 \citep{2022Sci...378..538I}, 3C 120 is another neutrino emitting candidate. However, origins of the neutrino production of these two sources are likely different. As a radio quiet source, neutrinos of NGC 1068 are proposed to be from outflow-ISM interaction roughly 50~pc away from the central engine \citep{2023ApJ...956....8F}, or at a place near the SMBH \citep{2023ApJ...954L..49A}. Note that the averaged $\gamma$-ray luminosity of 3C 120 is two orders of magnitude higher than that of NGC 1068 (see Figure~\ref{GLN}), and $\gamma$-ray emissions of 3C 120 are widely accepted to be from the relativistic jet \citep[e.g.,][]{2010ApJ...720..912A,2011ApJ...740...29K,2016MNRAS.458.2360J}. The temporal coincidences between the arrival of the neutrino and the $\gamma$-ray flare further strengthen the connection between the neutrino and the jet. Results from theoretical studies support that the relativistic jets in radio galaxies are possible neutrino contributors \citep{2014PhRvD..89l3005B,2017JCAP...12..017B}. Besides NGC 1068, recent studies have provided accumulating observational evidences that nearby radio-quiet Seyfert galaxies are potential neutrino sources. Neutrino excesses from the direction of NGC 4151 and CGCG 420-015 are reported, at post-trial significances of nearly 3$\sigma$ \citep{2025ApJ...988..141A,2025ApJ...981..131A}, although for the former blazars in the same direction may contribute the neutrino radiations as well \citep{2025A&A...694A.203O}. Analyses of publicly available ten-year IceCube dataset also suggest that NGC 3079 is likely another neutrino-emitting source \citep{2024PhRvL.132j1002N}. Interestingly, two individual neutrino events, IC-220424A and IC-230416A, are co-spatial with NGC 7469 \citep{2025ApJ...981..103S}. These evidences are not only important to explore the origin of the extragalactic neutrinos, but also shed a new light on the innermost region of the SMBH \citep[e.g.,][]{2024ApJ...974...75F}.

It is also intriguing to compare 3C 120 with blazars that both spatially and temporally coincide with the incoming neutrinos \citep[e.g.,][]{2018Sci...361.1378I,2019ApJ...880..103G,2023MNRAS.519.1396S,2022ApJ...932L..25L,2024ApJ...965L...2J,2016NatPh..12..807K,2020A&A...640L...4G}. As shown in Figure~\ref{GLN}, the averaged $\gamma$-ray luminosity of 3C 120 ($<~10^{44}$ erg $\rm s^{-1}$) is generally two orders of magnitudes lower than that of the blazars $\sim 10^{46}$ erg $\rm s^{-1}$. Due to the relatively large inclination jet angle, the Doppler boost effect of 3C 120 is likely suppressed. As the first neutrino emitting radio galaxy candidate, 3C 120 serves as a unique target, which offers a different perspective than blazars for approaching the production of neutrinos. In fact, 3C 120 fills in the blank between the blazars and NGC 1068. Moreover, analyses of the IceCube data suggest that luminous sources (i.e., blazars) are unlikely the dominant population for extragalactic neutrinos \citep{2017ApJ...835...45A}. On the other hand, relatively less powerful sources, such as nearby radio galaxies, starbursts, as well as galaxy clusters and groups, are likely preferred, and a detection of the few brightest objects is enabled \citep{2016PhRvD..94j3006M}. Our study on 3C 120 indeed encourages such a proposition. Future more cases like 3C 120 would be crucial to a comprehensive understanding on the source of the IceCube diffuse neutrinos.

In summary, based on the multi-wavelength observations of 3C 120, we suggest that it is likely associated with the neutrino event IC-180213A. 3C 120 is identified as the unique co-spatial $\gamma$-ray source at the time of the neutrino’s arrival. Moreover, a prominent $\gamma$-ray flare that is second strongest one among the entire 16-year emerges then. Correlated optical brightenings in the V and $g$-bands observed by ASAS-SN have also detected. 
Monte Carlo simulations give a probability of $\sim 0.04$ for a chance association. Although the estimated significance is a posteriori and it should be treated with caution, a potential link between 3C 120 and the neutrino is suggested. It is the first radio galaxy proposed as a neutrino emitter, filling in the blank between the blazars and the nearby radio quiet seyfert galaxies, like NGC 1068. Theoretical constraints on the jet properties of 3C 120 have been investigated and comparisons with other neutrino emitting candidates have been performed.

\begin{acknowledgements}
We appreciate the instructive suggestions from the anonymous referee that led to a substantial improvement of this work. We appreciate Dr. A.~B.~Pushkarev for sharing their adaptive-binned light curve data of 3C~120. This research has made use of data obtained from the High Energy Astrophysics Science Archive Research Center (HEASARC), provided by NASA's Goddard Space Flight Center. This research has made use of data from the ASAS-SN project, which is supported by the Ohio State University and operated by the Ohio State Astronomy Department. This research has made use of the NASA/IPAC Infrared Science Archive, which is funded by the NASA and operated by the California Institute of Technology. This research uses data products from the Wide-field Infrared Survey Explorer, which is a joint project of the University of California, Los Angeles, and the Jet Propulsion Laboratory/California Institute of Technology, funded by the National Aeronautics and Space Administration. This research also makes use of data products from NEOWISE-R, which is a project of the Jet Propulsion Laboratory/California Institute of Technology, funded by the Planetary Science Division of the National Aeronautics and Space Administration. This research has made use of the RATAN-600 data, provided by the Special Astrophysical Observatory of the Russian Academy of Sciences (SAO RAS). 

This work was supported in part by the NSFC under grants U2031120. This work was also supported in part by the Guizhou Provincial Science and Technology Projects (No. QKHFQ[2023]003 and No. QKHPTRC-ZDSYS[2023]003).

\end{acknowledgements}

\bibliographystyle{aa}
\bibliography{references}

\onecolumn

\end{document}